\documentclass[aps,prl,longbibliography,twocolumn,superscriptaddress,amsfont,graphicx,nofootinbib,preprintnumbers]{revtex4-1}%
\UseRawInputEncoding
\usepackage{color,graphicx,epsfig}
\usepackage{ifpdf}
\usepackage{amsmath}
\usepackage{bm}
\usepackage[english]{babel}
\usepackage{amssymb}
\usepackage{braket}
\usepackage{setspace}
\usepackage{epstopdf}
\allowdisplaybreaks[4]

\usepackage{hyperref}
\usepackage{enumerate}
\usepackage{lipsum}

\usepackage{slashed}

\usepackage{changes}

\begin{document}

\title{Probing Supernova Neutrino Boosted Dark Matter with Collective Excitation}%


\author{Jun-Wei Sun}
\affiliation{School of Physics, Yantai University, Yantai 264005, China}

\author{Lei Wu}
\thanks{Corresponding author}
\email{leiwu@njnu.edu.cn}
\affiliation{Department of Physics and Institute of Theoretical Physics, Nanjing Normal University, Nanjing, 210023, China}

\author{Yan-Hao Xu}
\thanks{Corresponding author}
\email{xuyanhao@ytu.edu.cn}
\affiliation{School of Physics, Yantai University, Yantai 264005, China}

\author{Bin Zhu}
\thanks{Corresponding author}
\email{zhubin@mail.nankai.edu.cn}
\affiliation{School of Physics, Yantai University, Yantai 264005, China}


\begin{abstract}

We explore the supernova neutrino-boosted dark matter (SN$\nu$BDM) and its direct detection. During core-collapse supernovae, an abundance of neutrinos are emitted. These supernova neutrinos can transfer their kinetic energy to the light dark matter via their interactions, and thus are detected in the neutrino and dark matter experiments. Due to the exponential suppression of the high-energy neutrino flux, the kinetic energy carried by a large portion of SN$\nu$BDM falls within the MeV range. This could potentially produce the collective excitation signals in the semiconductor detectors with the skipper-CCD. We calculate the plasmon excitation rate induced by SN$\nu$BDM and derive the exclusion limits. In contrast with conventional neutrino and dark matter direct detection experiments, our results present a significant enhancement in sensitivity for the sub-MeV dark matter.
\end{abstract}


\maketitle

\section{Introduction}
The nature of dark matter (DM) remains one of the most pressing open questions in modern physics. While gravitational evidence for dark matter is well-established, its non-gravitational interactions and intrinsic particle properties remain a mystery. Over the past few decades, the pursuit of Weakly Interacting Massive Particles (WIMPs) has been at the forefront of dark matter research. However, the absence of positive signals~~\cite{CDEX:2020tkb, PandaX-4T:2021bab, LZ:2022ufs, XENON:2023sxq} has shifted interest toward sub-GeV DM, which presents new opportunities for detection through unique signatures, such as electron recoils~~\cite{Essig:2011nj, Essig:2015cda, Essig:2017kqs, Emken:2019tni, Trickle:2020oki, Andersson:2020uwc, Vahsen:2021gnb, Su:2021jvk, Catena:2021qsr, Elor:2021swj, Du:2022dxf, Chen:2022pyd, Catena:2022fnk,  Su:2022wpj, Wang:2023xgm,Li:2022jxo,Bhattiprolu:2023akk,Su:2023zgr, Liang:2024ecw,Liang:2024lkk,He:2024hkr}, phonon excitation~\cite{Knapen:2017ekk,Raya-Moreno:2023fiw, Mitridate:2023izi,Kahn:2020fef, Guo:2024sqh} or the Migdal effect~\cite{Ibe:2017yqa, Dolan:2017xbu, Baxter:2019pnz, Bell:2019egg,Liang:2019nnx, Essig:2019xkx,GrillidiCortona:2020owp, Liu:2020pat, Liang:2020ryg, Flambaum:2020xxo, Acevedo:2021kly, Bell:2021zkr, Bell:2021ihi, Cox:2022ekg, Li:2022acp}.

The challenge of detecting low-energy nuclear recoils from sub-GeV DM is demonstrated by the typically low kinetic energy of dark matter particles in the galactic halo. To overcome this limitation, additional kinetic energy is required to elevate these particles' velocities, enabling them to surpass detector thresholds. One promising avenue for enhancing the kinetic energy of DM particles is through interactions with high-energy cosmic-rays~\cite{Bringmann:2018cvk, Ema:2018bih, Cappiello:2018hsu, Dent:2019krz, Wang:2019jtk, Alvey:2019zaa, Su:2020zny, Ge:2020yuf, Cao:2020bwd,Du:2020ybt,Jho:2020sku,Bloch:2020uzh,Lei:2020mii,Guo:2020oum,Dent:2020syp, Guo:2020oum, Wang:2021oha,Xia:2021vbz, Bell:2021xff, Wang:2021jic, Feng:2021hyz, Li:2022dqa,Bhowmick:2022zkj,Xia:2024ryt,Guha:2024mjr,Cappiello:2024acu,Xia:2024ryt, Ge:2024cto}. However, we argue that dark matter boosted by neutrinos presents a compelling and attractive concept. Neutrinos, due to their tiny masses and extremely weak interactions, are able to propagate through the vast majority of cosmic structures without significant scattering. This unique property allows neutrinos to interact with DM particles in a relatively direct and unperturbed manner, largely unaffected by the presence of other matter in the universe thus having small uncertainty in direct detection.

\begin{figure}
    \centering
    \includegraphics[width=0.48\textwidth]{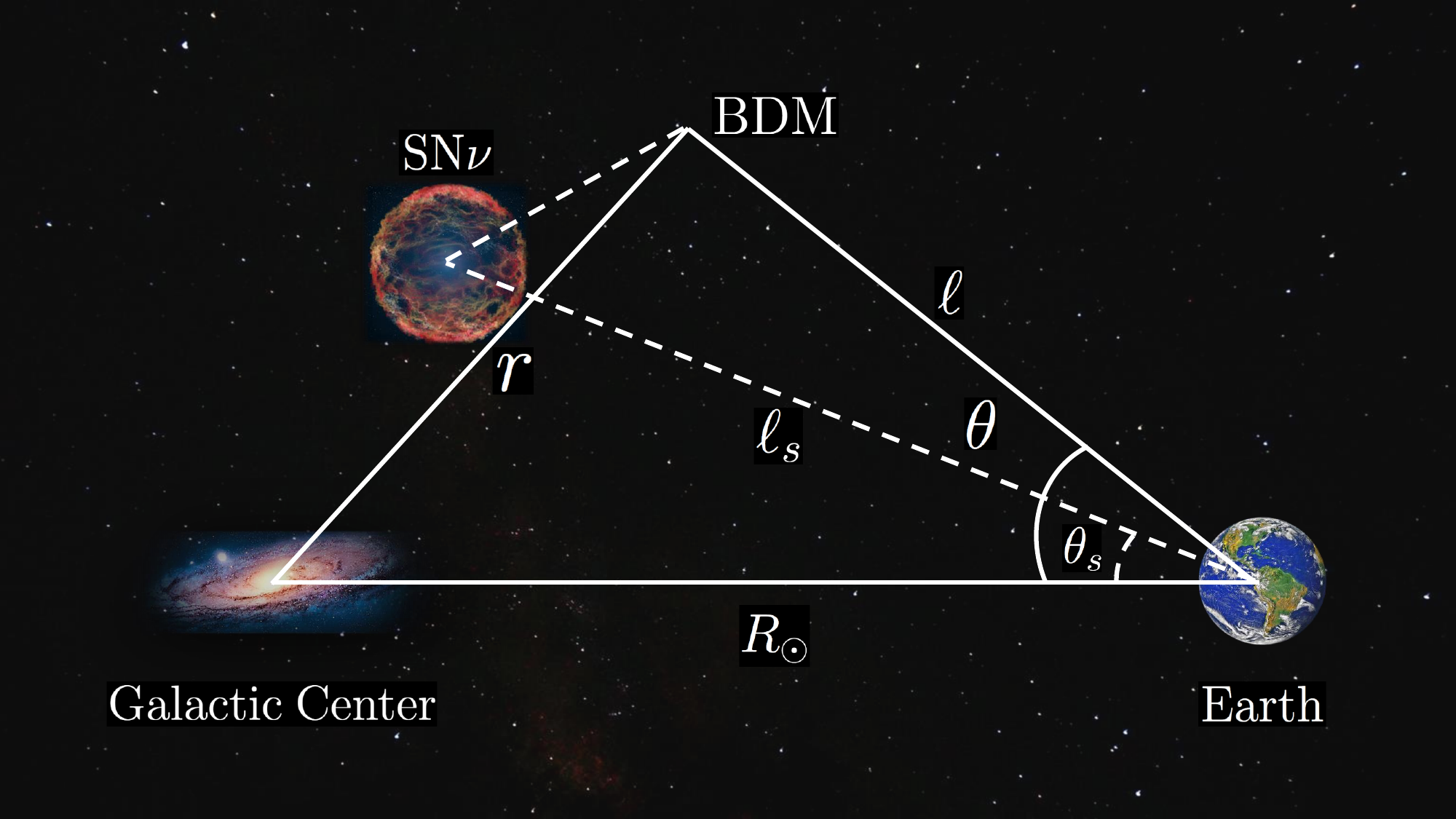}
\caption{{The neutrino flux images produced by  CC SNe at a specific location $\ell$, in which we establish the coordinate system of CC SNe.}}
    \label{Fig.1}
\end{figure}

In this work, we propose that galactic Core-Collapse supernova neutrinos (CC SNe) can significantly enhance the sensitivity of light dark matter detection, since CC SNe are amongst the brightest astrophysical sources of neutrinos. By transferring kinetic energy to dark matter particles, supernova neutrinos can facilitate their detection by enabling them to more effectively overcome energy thresholds~\cite{Jho:2021rmn, Das:2021lcr, Bardhan:2022bdg, Cline:2022qld, Lin:2022dbl, Lin:2023nsm, Ghosh:2024dqw, Lin:2024vzy}. Different from the previous works, we consider the contributions from galactic CC SNe, instead of distant CC SNe throughout the Universe, the so-called Diffuse SN Background (DSNB). This allows us to derive an analytical expression for the neutrino flux that varies across different locations within the Milky Way. Such a detailed treatment is crucial, as it enables the inclusion of dark matter scattering at any location in the galaxy, enhancing the potential for detection. On the other hand, when dark matter is accelerated by supernova neutrinos, namely SN$\nu$BDM, the dark matter velocity will be on the order of \(10^{-2} c\) because of the dynamics of SN$\nu$BDM. This may lead to the plasmon excitation in the semiconductor system~\cite{Liang:2024xcx, Essig:2024ebk}, including SENSEI \cite{SENSEI:2023zdf}, DAMIC \cite{Castello-Mor:2020jhd}, CDEX \cite{CDEX:2022kcd},EDELWEISS \cite{EDELWEISS:2020fxc}, and SuperCDMS \cite{SuperCDMS:2020ymb}. 

Utilizing the dataset obtained from the silicon detector SENSEI at SNOLAB~\cite{SENSEI:2023zdf}, we derive a novel constraint on the scattering cross section between DM and electrons. This new bound is found to be approximately $3$ to $4$ orders of magnitude more stringent than the limits previously established by the Super-Kamiokande experiment~\cite{Super-Kamiokande:2023jbt}, specifically for DM masses in the range of 1 keV to 1 MeV. Our analysis highlights the potential of utilizing these collective modes to improve the observability of dark matter interactions, offering a promising direction for future experimental efforts.

\section{Characteristics of Galactic Supernova Neutrinos} 
The time-averaged neutrino flux (particle number per unit of time, area, and energy) from CC SNe at the position $\bm{\ell}$, can be expressed as
\begin{equation}
    \frac{d\phi_\nu (\bm{\ell})}{dE_\nu} = \dot{N}_{\text{CC}}^{\text{gal}} \frac{dN_\nu}{dE_\nu} \int \frac{f(\bm{\ell}_s)}{4\pi |\bm{\ell}-\bm{\ell}_s|^{^2}} d^3\ell_s,
    \label{eq:neutrino_flux}
\end{equation}
where ${dN_\nu}/{dE_\nu}$ is the neutrino spectrum, $\dot{N}_{\text{CC}}^{\text{gal}}$ denotes the galactic core-collapse supernova rate, taken as $2.3 \times 10^{-2} \text{yr}^{-1}$~\cite{Li_2011}. In Fig.~\ref{Fig.1}, we illustrate the geometry of supernova neutrinos and their potential scattering with dark matter. The Earth is chosen as the origin of the coordinate system, with the Galactic Center located at a distance $R_{\odot}=8.5~\text{kpc}$ from the Earth. $f(\bm{\ell}_s)$ represents the spatial probability density function for a CC SNe occurring at position $\bm{\ell}_s$, while the scattering event between the neutrinos and dark matter takes place at the vector $\bm{\ell}$. The importance of considering the probability of CC SNe across entire galaxies is evident in Fig.~\ref{Fig.1}, where CC SNe can interact with the DM halo at various locations, transferring kinetic energy to DM. To account for all potential sources of DM flux, it is essential to specify the CC SNe flux at different locations throughout the galaxy. {It should be emphasized that only the Supernova at a particular position is shown in Fig.~\ref{Fig.1}, and the actual supernova distribution is given by the following Eq.~\eqref{eq:f(l)}}.

According to Ref.~\cite{Adams:2013ana}, the spatial distribution of CC SNe can be modeled as a double-exponential function in galactocentric cylindrical coordinates, $\rho(R,z)\sim \exp(-R/R_d-|z|/z_H)$, where $R$ is the Galactocentric radius and $z$ is the height above
the Galactic plane, $R_d$=2.9~\text{kpc} is scale length parameter and $z_H=95~\text{pc}$ is scale high parameter. $f(\bm{\ell}_s)$ can be obtained by performing a coordinate transformation from the GC to the Earth. For spherical coordinates $(\ell_s,\theta_s,\phi_s)$, the $f(\bm{\ell}_s)$ can be written as
\begin{equation}
    f(\bm{\ell}_s)=\frac{1}{\mathcal{N}}\exp \left[-\frac{R(\ell_s,\theta_s,\phi_s)}{R_d}-\frac{|z(\ell_s,\theta_s,\phi_s)|}{ z_H}  \right],
    \label{eq:f(l)}
\end{equation}
where $\mathcal{N}$ denotes the normalization factor to ensure that the integral of $f(\bm{\ell}_s)$ over all space is one, $\int f(\bm{\ell}_s)d^3\ell_s=1$. The geometric relationship between cylindrical and spherical coordinates is as follows
\begin{equation}
\begin{aligned}
    R&=\sqrt{R_\odot^2+\ell_s^2-\ell_s^2\sin^2\theta_s \sin^2\phi_s-2\ell_s R_\odot \cos\theta_s},\\
    z&=\ell_s \sin\theta_s \sin\phi_s.
\end{aligned}
\end{equation}

The neutrino spectrum (particle number per unit of energy) from a supernova can be approximated by a pinched Fermi-Dirac distribution~\cite{Keil:2002in}. For a certain type of neutrino, the spectrum can be written as
\begin{equation}
     \frac{dN}{dE_{\nu_i}} = \frac{E_{\text{tot}}^\nu (1 + \alpha)^{1+\alpha}}{\Gamma(1 + \alpha)} \frac{E_\nu^\alpha}{\langle E_\nu \rangle^{2+\alpha}} \exp\left[-(1+\alpha) \frac{E_\nu}{\langle E_\nu \rangle}\right],
    \label{eq:neutrino_spectrum}
\end{equation}
where $E_{\text{tot}}^\nu$ is the total energy radiated in neutrinos, $\langle E_\nu \rangle$ is the average neutrino energy, and $\alpha$ is the spectral shape parameter. In table~\ref{tab:neutrino_par} we show the values for different neutrinos adopted in this work. The total neutrino flux is the sum of all neutrino flavors,
\begin{equation}
   \frac{dN}{dE_\nu}=\frac{dN}{dE_{\nu_e}}+
   \frac{dN}{dE_{\nu_{\bar{e}}}}+4\frac{dN}{dE_{\nu_x}},
\end{equation}
where $\nu_x$ denotes four types of neutrinos, which are $\nu_\mu, \nu_{\bar{\mu}}, \nu_\tau, \nu_{\bar{\tau}}$ respectively.
\begin{table}[t]
    \centering
    \setlength{\tabcolsep}{10 pt}
    \begin{tabular}{cccc}
    \hline\hline
       $\nu$  & $E_{\text{tot}}^\nu ~(\text{erg})$  &$\langle E_\nu \rangle ~(\text{MeV})$  & $\alpha$\\
       \hline
       $\nu_e$ & $6\times 10^{52}$ & 13.3 & 3.0\\
       \hline
       $\nu_{\bar{e}}$ & $4.3\times 10^{52}$ & 14.6 & 3.3\\  
       \hline
       $\nu_x$  &  $2\times 10^{52}$ & 16 & 3\\       
       \hline\hline
    \end{tabular}
    \caption{Values of parameters used in \eqref{eq:neutrino_spectrum} for electron neutrinos $\nu_e$, anti-electron neutrinos $\nu_{\bar{e}}$, and other neutrinos $\nu_x \equiv \{ \nu_\mu, \nu_{\bar{\mu}}, \nu_\tau, \nu_{\bar{\tau}} \} $, taken from Ref.~\cite{Horiuchi:2017qja}.}
    \label{tab:neutrino_par}
\end{table}

Note that there is a particular case worth commenting on. If the supernova is a point source only exists at the GC, the probability density function $f(\bm{\ell}_s)$ can be simplified as a 3-dimensional $\delta$ function, $f(\bm{\ell}_s)=\delta^3(\bm{\ell}_s-\bm{\ell}_{GC})$, where $\bm{\ell}_{GC}$ denotes the position vector of the GC. In addition, if we assume one supernova explosion with a duration time of $\tau$ and the neutrino propagates with the velocity of light $c$, the time-averaged flux in a spherical shell with a thickness of $c\tau$ becomes:
\begin{equation}
    \frac{d\phi_\nu}{dE_\nu} = \frac{1}{\tau} \frac{dN_\nu}{dE_\nu} \frac{1}{4\pi |\bm{\ell}-\bm{\ell}_{GC}|^{^2}}.
\end{equation}
This is exactly the flux adopted in Ref.~\cite{Lin:2022dbl, Lin:2023nsm}. When studying supernova neutrinos, we prioritize galactic supernova neutrinos over the diffuse supernova neutrino background (DSNB) for several reasons. First, the flux of galactic supernova neutrinos at Earth is significantly higher than that of the DSNB due to cosmic distance dilution affecting the latter. This higher flux allows for more precise measurements and better constraints, which are essential for investigating CC SNe.

\section{Benchmark Model}
Before determining the incident flux and the event rate for the targeted particle after entering the detector, we must identify the differential scattering cross section of the dark matter-Standard Model particle interaction.  Taking the interaction between dark matter and leptons as an example, we consider the coupling of a singlet dark matter fermion $\chi$ with a vector boson $Z^{\prime}_{\mu}$. Such an effective Lagrangian is leptophilic dark matter~\cite{PhysRevD.77.023506, Falkowski:2009yz, PhysRevD.79.083528, PhysRevD.90.015011, GonzalezMacias:2015rxl, Blennow:2019fhy, FileviezPerez:2019cyn},
\begin{equation}
\mathcal{L} \supset  g_{\nu} \Bar{\nu} \gamma^{\mu} \nu Z^{\prime}_{\mu} + g_{e} \Bar{e} \gamma^{\mu} e Z'_{\mu}+g_{\chi} \Bar{\chi} \gamma^{\mu} \chi Z^{\prime}_{\mu}.
\label{eq:lagrangian_nu_DM}
\end{equation}
Here, $g_{\nu}$, $g_{e}$, and $g_{\chi}$ denote the coupling constants corresponding to neutrinos, electrons, and dark matter, respectively. In a supernova, the $Z^{\prime}$ can be generated through the coalescence of (anti-)neutrinos within the stellar core. These mediators may freely escape the supernova and subsequently decay into a pair of (anti)neutrinos, thereby producing an (anti)neutrino flux that extends beyond the Standard Model~\cite{Magill:2018jla, Brdar:2020quo, Akita:2022etk, Fiorillo:2022cdq}. Consequently, it is essential to account for this back-reaction type of neutrino flux to maintain consistency in our analysis. However, it is important to note that, to significantly excite the plasmon in silicon detectors, the mediator $Z^{\prime}$ should ideally light~\cite{Liang:2024xcx}, which renders decay production highly suppressed. Even when considering a heavier mediator, such as $200\ \mathrm{MeV}$, the modifications to the neutrino flux occur at the $100\ \mathrm{MeV}$ scale~\cite{Akita:2023iwq, Lazar:2024ovc}, which does not contribute meaningfully to plasmon excitation. Therefore, we can safely neglect the influence of the mediator on the neutrino flux.

The dynamics of the scattering process remain Lorentz invariant regardless of events occurring on Earth or in the halo. However, we primarily focus on the laboratory frame, which is more suitable for data analysis. Thus, we divide the cross section into two types with different kinematics: one corresponding to the interaction between supernova neutrinos (cosmic electrons) and dark matter in the galaxy, and the other involving boosted dark matter interacting with electrons in the laboratory.

This interaction Eq.~\eqref{eq:lagrangian_nu_DM} manifests as a scattering process mediated by the $t$-channel process $\nu (e) + \chi \rightarrow \nu (e) + \chi$, facilitating the transfer of kinetic energy to halo dark matter from supernova neutrinos or cosmic electrons. The resulting scattering amplitude for the neutrino-dark matter interaction is expressed as follows:
\begin{equation}
|\mathcal{M}|_{\chi \nu}^{2} = \frac{2 g_{\nu}^{2} g_{\chi}^{2}}{(m_{Z'}^{2} - t)^{2}} \left[ 2(m_{\chi}^{2} - s)^{2} + 2st + t^{2} \right],
\label{eq:Msquared_nu_DM}
\end{equation}
where $m_{Z^{\prime}}$ signifies the mass of the exchanged vector mediator, and $t$ and $s$ are the Mandelstam variables, while $m_{\chi}$ represents the mass of the dark matter. For the interaction of electrons with dark matter, we also observe the same situation, with the scattering amplitudes as follows:
\begin{equation}
|\mathcal{M}|_{\chi e}^{2} = \frac{2 g_{e}^{2} g_{\chi}^{2}}{(m_{Z'}^{2} - t)^{2}} \left[ 2(m_{e}^{2}+m_{\chi}^{2} - s)^{2} + 2st + t^{2} \right].
\label{eq:Msquared_e_DM}
\end{equation}
The differential scattering cross-section related to the kinetic energy distribution of dark matter can be derived,
\begin{equation}
 \frac{d\sigma}{d T_{\chi}}=\frac{d\sigma}{dt} \left|\frac{dt}{dT_{\chi}}\right|=\frac{2m_{\chi}}{64 \pi s |\mathbf{p}|^{2}} |\mathcal{M}|^{2}.
\label{eq:cs}
\end{equation}
In the context of neutrino (electron)-dark matter interactions, termed boosted dark matter, the ultimate formulations for the differential scattering cross-section are defined as 
\begin{equation}
\begin{aligned}
\frac{d \sigma_{\chi \nu}}{d T_{\chi}} &= \frac{g_{\nu}^{2} g_{\chi}^{2} m_{\chi}[2E_{\nu}^{2}-(2E_{\nu}+m_{\chi}) T_{\chi}+T_{\chi}^{2}]}{4\pi E_{\nu}^{2} (m_{Z'}^{2}+2 m_{\chi} T_{\chi})^{2}},\\
\frac{d \sigma_{\chi e}}{d T_{\chi}} &= \frac{g_{e}^{2} g_{\chi}^{2} [2 E_{e}'^{2} m_{\chi}-(m_{e}^{2}+m_{\chi}(2 E_{e}'+m_{\chi})) T_{\chi}+m_{\chi}T_{\chi}^{2}]}{4\pi (E_{e}'^{2}-m_{e}^{2}) (m_{Z'}^{2}+2 m_{\chi} T_{\chi})^{2}}. 
\end{aligned}
\label{eq:cs nu DM}
\end{equation}

where $E_{\nu}$ is the neutrino kinetic energy and $E_{e}'$ is the total electron enery

Unlike cosmic electron scattering, semiconductor detectors consider the interaction between boosted dark matter and electrons by involving the many-body effect~\cite{ Hochberg:2017wce, Knapen:2017ekk, Hochberg:2016sqx, Hochberg:2016ntt, Bloch:2016sjj, Griffin:2018bjn, Trickle:2019nya, Griffin:2019mvc, Geilhufe:2019ndy, Griffin:2020lgd, Coskuner:2019odd, Cavoto:2017otc, Trickle:2019ovy, Ge:2017mcq, Hochberg:2021pkt, Roberts:2016xfw, Kahn:2021ttr, Blanco:2019lrf, Knapen:2021run, Griffin:2021znd, Ramanathan:2020fwm,  Kozaczuk:2020uzb, Ge:2022ius, Lasenby:2021wsc, Liang:2018bdb, Blanco:2022cel, Catena:2023qkj, Catena:2023awl, Gu:2022vgb}.  In the context of low-energy electronic excitation, the behavior of electrons in semiconductors is non-relativistic, allowing the corresponding Lagrangian to simplify into the form of the electron number density.
\begin{equation}
\mathcal{L}_{e}^{\mathrm{eff}} \supset g_{e}A_{0}^{\prime}\psi_{e}^{*}\psi_{e} + \frac{ig_{e}}{2m_{e}}\mathbf{A}^{\prime}\cdot\left(\psi_{e}^{*}\overrightarrow{\nabla}\psi_{e} - \psi_{e}^{*}\overleftarrow{\nabla}\psi_{e}\right) + \cdots,
\end{equation}
where $\psi_{e}$ denotes the nonrelativistic electron wavefunction and $m_{e}$ is the electron mass. For our analysis, only the first term, $g_{e}A_{0}^{\prime}\psi_{e}^{*}\psi_{e}$, is relevant, as the second term (representing the electric current) and higher-order contributions are suppressed by the electron mass $m_{e}$ in the context of the electron bound state wavefunction. As a consequence, the scattering amplitude can be expressed as
\begin{equation}
i \mathcal{M} = \Bar{u}_{\chi}' \gamma^{0} u_{\chi} \frac{i g_{e} g_{\chi}}{Q^{2} - E_{e}^{2} + m_{Z'}^{2}} \times \langle i|e^{i \mathbf{Q} \cdot \hat{\mathbf{x}}}|f \rangle,
\label{eq:M_DM_e}
\end{equation}
where $Q$ is the transferred momentum, $E_{e}$ is the energy transferred to the electron, and $\Bar{u}_{\chi}'$ and $u_{\chi}$ are the spinors of the initial and final dark matter states, respectively. Summing over the initial state spins and averaging over the final state spins, and taking into account the thermal distribution $\mathcal{P}_{i}$ of the initial electronic states $|i \rangle$ in the semiconductor, the square of the final scattering amplitude is expressed as
\begin{widetext}
\begin{equation}
|\mathcal{M}|^{2} = \frac{g_{e}^{2} g_{\chi}^{2}}{(Q^{2} - E_{e}^{2} + m_{Z'}^{2})^{2}} \frac{1}{2} \sum_{\Bar{u}' u} \Bar{u}'_{\chi} \gamma^{0} u_{\chi} \Bar{u}_{\chi} \gamma^{0} u'_{\chi} \sum_{i,f} \mathcal{P}_{i} \langle i|e^{-i \mathbf{Q} \cdot \hat{\mathbf{x}}}|f \rangle \langle f|e^{i \mathbf{Q} \cdot \hat{\mathbf{x}}}|i \rangle.
\label{eq:Msqrt_DM_e}
\end{equation}
\end{widetext}
Finally, the differential scattering cross section of dark matter interacting with electrons in a silicon semiconductor detector is expressed in terms of the reference cross section as
\begin{widetext}
\begin{equation}
\frac{d \sigma_{\chi e}^{\mathrm{Si}}}{d E_{e}}= 
2 N_{\mathrm{cell}} V_{\mathrm{cell}} \frac{\Bar{\sigma}_{\chi e} \pi}{\mu_{\chi e}^{2}} \int \frac{d \mathbf{Q}^{3}}{(2 \pi)^{3}} \frac{Q^{2}}{4 \pi \alpha}\frac{(2E_{\chi} - E_{e})^{2} - Q^{2}}{4 p_{\chi}^{2}} F_{\mathrm{DM}}(Q)^2 \mathrm{Im}\left[\frac{-1}{\epsilon(\mathbf{Q},E_{e})}\right],
\label{eq:cs_DM_e}
\end{equation}
\end{widetext}
where $N_{\mathrm{cell}}$ is the number of unit cells in the semiconductor, $V_{\mathrm{cell}}$ is the volume per unit cell, and $\mu_{\chi e}$ is the reduced mass of dark matter and electrons. Here, $\alpha$ is the fine structure constant, and $E_{\chi}$ and $p_{\chi}$ correspond to the total energy and momentum of dark matter, respectively. In Eq.~\eqref{eq:cs_DM_e}, $\mathrm{Im}(\epsilon^{-1}(\mathbf{Q},\omega))$ is the energy loss function, which can be obtained by simplifying $\sum \mathcal{P}_{i} \langle i|e^{-i \mathbf{Q} \cdot \hat{\mathbf{x}}}|f \rangle \langle f|e^{i \mathbf{Q} \cdot \hat{\mathbf{x}}}|i \rangle$. Thus, the dielectric function describes the hopping of electrons in a semiconductor. The reference cross section, $\Bar{\sigma}_{\chi e}$, is defined as
\begin{equation}
\Bar{\sigma}_{\chi e} = \frac{\mu_{\chi e}^{2}}{\pi} \left(\frac{g_{\chi} g_{e}}{\alpha^{2} m_{e}^{2} + m_{Z'}^{2}}\right)^{2},
\label{eq:reference_cs}
\end{equation}
and the dark matter form factor is defined as
\begin{equation}
    F_{\mathrm{DM}}(Q)^2=\left( \frac{\alpha^{2} m_{e}^{2} + m_{Z'}^{2}}{Q^{2} - E_{e}^{2} + m_{Z'}^{2}}\right)^{2}.
\end{equation}

When the mediator mass is much heavier than the transferred momentum \( Q \), the DM form factor satisfies \( F_{\mathrm{DM}}(Q)^2 = 1 \), which is referred to as the heavy mediator scenario. In contrast, in the limit where \( m_{Z}^{\prime} \ll Q \), the DM form factor becomes \( \left( {\alpha m_e}/{Q} \right)^4 \), known as the light mediator scenario. According to Ref.~\cite{Liang:2024xcx}, the plasmon enhancement is only significant in the light mediator scenario; therefore, we consider only the light mediator throughout this paper.

\section{Supernova Neutrino Boosted Dark Matter}
\label{Sec.Supernova Neutrino Boosted Dark Matter}
Notably, in most previous studies on neutrino-boosted dark matter~\cite{Jho:2021rmn, Das:2021lcr, Bardhan:2022bdg, Cline:2022qld, Lin:2022dbl, Lin:2023nsm, Ghosh:2024dqw}, a local (at $\bm{\ell}=0$) neutrino flux is adopted, neglecting spatial inhomogeneity. To elucidate the impact of neutrino inhomogeneity on the boosted dark matter flux, we can extract the position-dependent term from the neutrino flux expression Eq.~\eqref{eq:neutrino_flux} and define it as a factor 
\begin{equation}
K(\bm{\ell}) = \int \frac{f(\bm{\ell}_s)}{4\pi |\bm{\ell}-\bm{\ell}_s|^{2}} d^3\ell_s.
\end{equation}
Then, the flux of SN$\nu$BDM at the Earth's surface, observed from a specific direction, is expressed as
\begin{equation}
\frac{d\phi_\chi}{dT_\chi d\Omega} = D(\hat{\bm{\ell}}) \frac{\rho_{\chi}^{\rm{loc}}}{m_\chi} \int_{E_\nu^{\min}} \frac{d\sigma_{\chi \nu}}{dT_\chi} \frac{d\phi_{\nu}^{\rm{loc}}}{dE_\nu} dE_\nu,
\label{eq:bdm_flux_D}
\end{equation}
where 
\begin{equation}
D(\hat{\bm{\ell}}) = \frac{1}{4\pi} \int_{\mathrm{l.o.s.}} \frac{K(\bm{\ell})}{K(0)} \frac{\rho_\chi(\bm{r})}{\rho_{\chi}^{\rm{loc}}} d\ell.
\end{equation}
Here, $d\phi_{\chi}^{\rm{loc}}/dE_\nu$ denotes the local neutrino flux near the Earth, and $E_{\nu}^{\min}$ represents the minimum neutrino energy required to produce a given dark matter kinetic energy $T_\chi$. We consider elastic scattering while neglecting the neutrino mass, yielding
\begin{equation}
E_{\nu}^{\min} = \frac{T_{\chi}}{2} + \frac{1}{2} \sqrt{T_{\chi}(2m_{\chi} + T_{\chi})}.
\label{eq:neutrino_Emin}
\end{equation}

The dark matter density $\rho_\chi(\bm{r})$ at position $\bm{r}$ is defined with the Galactic Center (GC) as the origin. We adopt the standard Navarro-Frenk-White (NFW) profile, given by
\begin{equation}
\rho_{\chi}(\bm{r}) = \rho_{\chi}^{\mathrm{loc}} \left(1 + \frac{R_\odot}{R_s}\right)^2 \frac{R_\odot}{|\bm{r}|} \left(1 + \frac{|\bm{r}|}{R_s}\right)^{-2},
\label{eq:dm_profile}
\end{equation}
where $R_s = 20~\text{kpc}$, $R_\odot = 8.5~\text{kpc}$, and $\rho_{\chi}^{\mathrm{loc}} = 0.43~\mathrm{GeV~cm^{-3}}$ is the local dark matter density and the geometric relationship between $\bm{r}$ and
$\bm{\ell}$ can be obtained through the cosine formula as shown in Fig.~\ref{Fig.1}
\begin{equation}
    r^2=R_{\odot}^2+\ell^2-2R_\odot \ell \cos{\theta}.
\end{equation}
We define an angle-dependent distance factor $D(\hat{\bm{\ell}})$, integrated along the line of sight (l.o.s.), which incorporates all position-dependent information.

When deriving the formulas, we assume a uniform time distribution of supernova (SN) neutrinos, similar to the distribution of cosmic rays in galaxies. This assumption is valid if dark matter (DM) scattered by SN neutrinos from different sources at different locations arrives at Earth without significant time delay. However, for a single SN-boosted DM event, the flux does not last indefinitely, meaning it is possible for multiple sources of SN neutrino-boosted DM to arrive at Earth at different times. The requirement that these fluxes cannot be separated by time is that the duration time $t_{\mathrm{van}}$ must be larger than $100~\mathrm{years}$. Generally, $t_{\mathrm{van}}$ is proportional to the DM mass, which would imply a minimal DM mass in the MeV range~\cite{Lin:2022dbl, Lin:2023nsm}. However, a careful observation reveals that the duration time is also inversely proportional to the DM kinetic energy~\cite{Lin:2023nsm}. Thus, for keV-scale kinetic energy, even keV-scale DM can have a sufficiently long duration time. Therefore, we can treat the DM flux as arriving at Earth simultaneously, allowing us to use our formula to derive the time-averaged flux.

\begin{figure}[t]
    \centering    \includegraphics[width=0.45\textwidth]{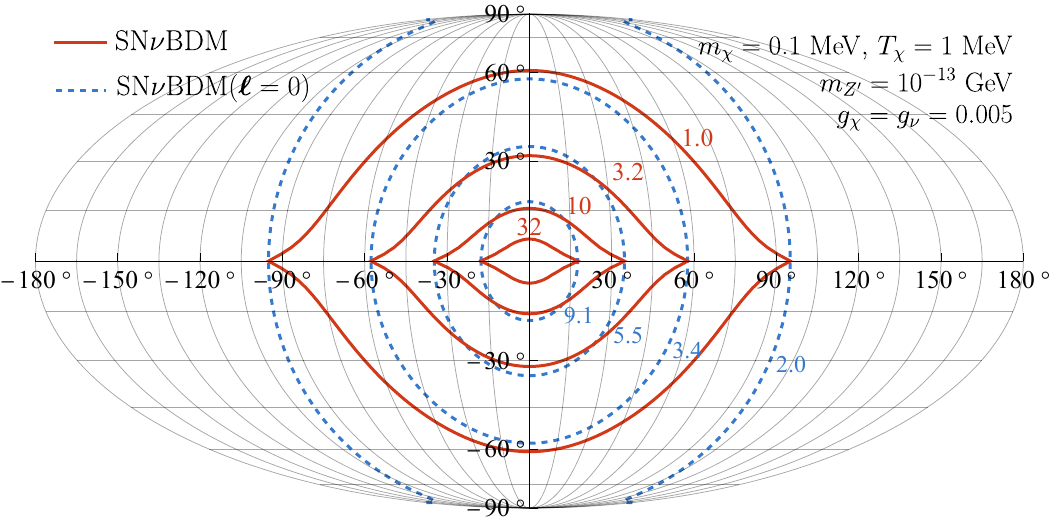}
\caption{The Contour of SN$\nu$BDM flux  $d\phi_\chi/dT_\chi d\Omega$ (in units of $\mathrm{GeV^{-1} cm^{-2} s^{-1} sr^{-1}}$) with the kinetic energy taken as $1~\mathrm{MeV}$ and $m_{\chi}=0.1~\mathrm{MeV}$. The coupling constants are set to $g_\chi=g_\nu=0.005$ and a light mediator $m_{Z'}=10^{-13}~\mathrm{GeV}$ is adopted. Red line represents the general result from \eqref{eq:bdm_flux_D} and the blue dashed shows the specific case from a local neutrino flux obtained by setting $\bm{\ell}=0$.}
    \label{fig:dmflux_contour}
\end{figure}

In Fig.~\ref{fig:dmflux_contour}, we present the contours of SN$\nu$BDM flux with the energy fixed at 1~MeV in the Galactic coordinate system. For comparison, we also include the results calculated using the local neutrino flux, obtained by setting $\bm{\ell}=0$ in Eq.~\eqref{eq:bdm_flux_D}. It is evident that the flux distribution with the general position-dependent neutrino flux is asymmetrically distributed around the Galactic Center (GC), consistent with our expectations, as the distribution of CC SNe is cylindrically symmetric, as implied in Eq.~\eqref{eq:f(l)}. In contrast, the result using the local neutrino flux exhibits azimuthal symmetry around the GC, as all angular information is derived from the spherically symmetric DM density profile $\rho_\chi(\bm{r})$. Notably, the SN$\nu$BDM flux is significantly higher in the direction of the GC compared to the local case, reaching approximately eight times greater. This disparity can be explained by Eq.~\eqref{eq:f(l)}, which indicates that the distribution of supernovae decreases exponentially from the GC to the surrounding region. This morphological feature closely resembles that observed in cosmic ray-boosted DM, which has been thoroughly studied in Ref.~\cite{Xia:2022tid}.

In direct detection, the total flux from all directions is commonly needed:
\begin{equation}
\frac{d\phi_\chi}{dT_\chi} = D_{\mathrm{eff}} \frac{\rho_{\chi}^{\rm{loc}}}{m_\chi} \int \frac{d\sigma_{\chi \nu}}{dT_\chi} \frac{d\phi_{\nu}^{\rm{loc}}}{dE_\nu} dE_\nu,
\label{Eq:BDMflux_Tchi}
\end{equation}
where the effective distance is defined as $D_{\rm{eff}} = \int D(\hat{\bm{\ell}}) d\Omega$. The value of $D_{\rm{eff}}$ varies due to the upper limit of the line-of-sight (l.o.s.) integration, which is the main source of uncertainty. In this work, we perform the full l.o.s. integration out to 30~kpc, which is sufficiently far since the integration converges rapidly due to the exponential suppression in $f(\bm{\ell})$. Our numerical calculation yields $D_\mathrm{eff} = 10.7~\rm{kpc}$, providing a relatively accurate value compared to neglecting the inhomogeneity of CC SNe, which results in $D_{\rm{eff}}=16.4~\rm{kpc}$. It is evident that the \(D_{\mathrm{eff}}\) calculated without considering inhomogeneity is approximately \(1.6\) times larger than our approach. This discrepancy results in significant uncertainty in quantifying the SN$\nu$BDM.

\begin{figure}
    \centering
    \includegraphics[width=0.48\textwidth]{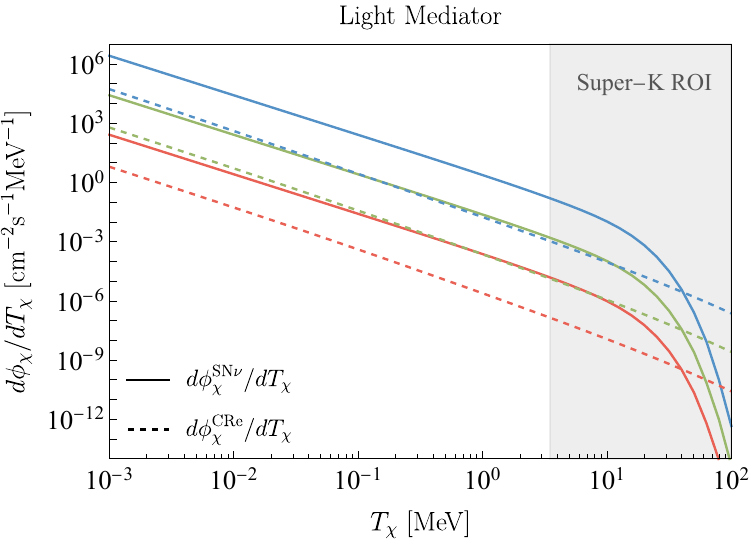}
\caption{Comparison of the SN$\nu$BDM flux with cosmic ray electron BMD flux after solid angle integration of SN$\nu$BDM flux with coupling constant $g_{\nu} = g_{\chi} = g_{e} = 0.005$. The blue, green and red colors respectively correspond to $m_{\chi}$ at $0.01~\mathrm{MeV}$, $0.1 ~\mathrm{MeV}$ and $1 ~\mathrm{MeV}$. The solid line indicates the SN$\nu$BDM flux, and the dashed line indicates the cosmic ray electron boosted dark matter flux. The gray area indicates the region of interest of the Super-Kamiokand experiment~\cite{Super-Kamiokande:2023jbt}.}
    \label{Fig.3}
\end{figure}

Fig.~\ref{Fig.3} illustrates the SN$\nu$BDM flux and cosmic electron boosted DM (CRDM) flux for $m_{\chi}$ equaling $0.01~\mathrm{MeV}$, $0.1 ~\mathrm{MeV}$ and $1 ~\mathrm{MeV}$, where the CRDM flux is calculated according to Ref.~\cite{Xia:2022tid} using the energy dependent cross section from Eq.~\eqref{eq:cs nu DM}. The flux exhibits a decreasing trend as the dark matter kinetic energy \( T_{\chi} \) increases since the differential scattering cross section is inversely proportional to $T_{\chi}$. 
Notably, the SN$\nu$BDM fluxes decline more rapidly when kinetic energy above 10~MeV, which originates from the exponential suppression term in the neutrino spectrum as shown in Eq.~\eqref{eq:neutrino_spectrum}. This rapidly declining property guarantees the better sensitivity of plasmon excitation than the conventional elastic scattering in neutrino detector, as the plasmon is only sensitive to the low energy processes. It is generally expected that neutrino detectors have an advantage due to their much larger exposure. However, for SN$\nu$BDM, the significant decline in the MeV range reduces the flux above the threshold of neutrino detectors like Super-Kamiokande~\cite{Super-Kamiokande:2023jbt}. Consequently, we anticipate achieving higher sensitivity than Super-Kamiokande~\cite{Super-Kamiokande:2023jbt} in detecting SN$\nu$BDM, which will be proven in the next section. 

It follows from Eq.~\eqref{Eq:BDMflux_Tchi} that the impact of the dark matter mass $m_{\chi}$ on the SN$\nu$BDM flux is determined by two parts: the dark matter number density $n_{\chi}=\rho_{\chi}/m_{\chi}$ and the scattering cross section $d\sigma_{\chi\nu}/dT_{\chi}$. Explicitly, the dark matter number density is inversely proportional to $m_{\chi}$. Given our choice of a light mediator, the scattering cross section also exhibits an inverse proportionality to $m_{\chi}$. Consequently, these two factors result in a $1/m_{\chi}^2$ scaling of the SN$\nu$BDM flux for varying $m_{\chi}$, explaining the observed hierarchy among different SN$\nu$BDM fluxes in Fig.~\ref{Fig.3}. A similar trend is observed for CRDM fluxes. Moreover, for a fixed dark matter mass  $m_{\chi}$, the CRDM flux is approximately $1.5$ orders of magnitude smaller than the SN$\nu$BDM flux at $T_{\chi} < 10~\mathrm{MeV}$, assuming identical coupling constants $g_{\nu} = g_{e} = g_{\chi} = 0.005$. This difference can be attributed to the interstellar spectrum of cosmic ray electrons, which, at MeV energies, is substantially reduced by $1.5$ orders of magnitude compared to the CC SNe flux.

\section{Plasmon-enhanced Event Rate and Numerical Results}
\begin{figure*}[ht]
    \centering
\includegraphics[width=8cm]{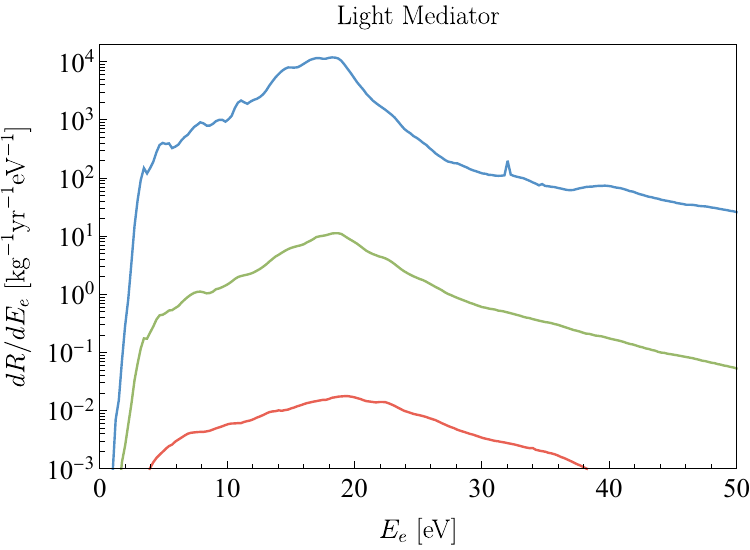}
\includegraphics[width=8cm]{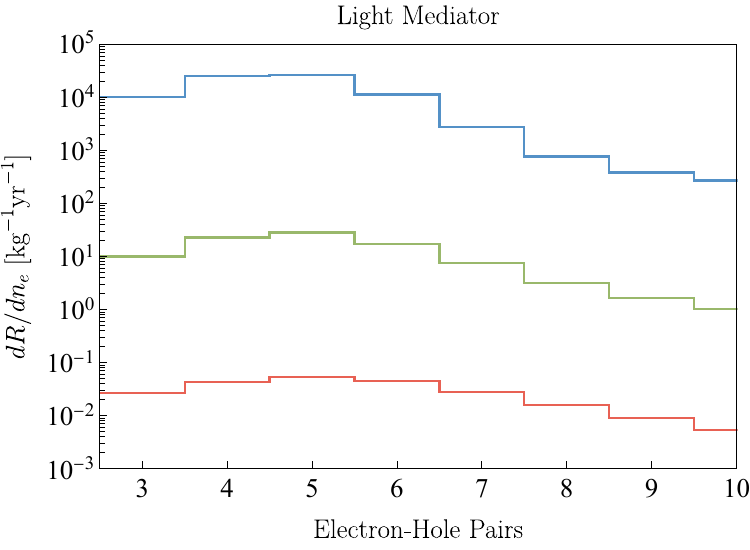}
\caption{Differential event rate of dark matter and electron scattering after considering the Plasmon-enhanced result. Where the red, green and blue solid lines respectively represent $m_{\chi}=0.01 \mathrm{MeV}$, $m_{\chi}=0.1 \mathrm{MeV}$ and $m_{\chi}=1 \mathrm{MeV}$.
\textbf{Left:} Differential event rate versus electron recoil energy $E_{e}$. \textbf{Right:} Differential event rates versus electron-hole pairs.}
    \label{Fig.4}
\end{figure*}

This section investigates the potential of plasmon resonance excitation induced by SN$\nu$BDM in semiconductor detectors, specifically the SENSEI at SNOLAB~\cite{SENSEI:2023zdf}. The cross-section for interactions between SN$\nu$BDM and electrons is significantly enhanced due to the resonant nature of plasmon excitation, potentially leading to a higher event rate than traditional detection approaches, as illustrated in Fig.~\ref{Fig.4}. The process is quantitatively described by the triple differential event rate
\begin{equation}
\begin{aligned}
\frac{dR}{dT_{\chi} dQ dE_{e}} &= \bar{\sigma}_{\chi e} \frac{N_{\mathrm{cell}} V_{\mathrm{cell}}}{8 \pi^{2} \alpha \mu_{\chi e}^{2}} \frac{((2E_{\chi} - E_{e})^{2} - Q^{2}) Q^{3}}{4 p_{\chi}^{2}} \\
&\left(\frac{\alpha^{2} m_{e}^{2} + m_{A'}^{2}}{Q^{2} - E_{e}^{2} + m_{A'}^{2}}\right)^{2} \mathrm{Im}\left[\frac{-1}{\epsilon(Q, E_{e})}\right] \frac{d\phi_{\chi}}{d T_{\chi}},
\end{aligned}
\label{func}
\end{equation}
where we consider the detector to be isotropic, allowing the transfer momentum \( \mathbf{Q} \) to be reduced to its magnitude \( Q \). The kinematic constraints require that the energy be limited to

\begin{equation}
\sqrt{(p_{\chi} + Q)^{2} + m_{\chi}^{2}} \geq E_{\chi} - E_{e} \geq \sqrt{(p_{\chi} - Q)^{2} + m_{\chi}^{2}}.
\label{Kinesiological constraint}
\end{equation}

With the kinematic limits of Eq.~\eqref{Kinesiological constraint}, we are able to obtain the limits $Q_{\mathrm{min}}$ and $Q_{\mathrm{max}}$. By integrating \( T_{\chi} \) and \( Q \) in Eq.~\eqref{func}, we obtain the differential event rate:

\begin{equation}
\frac{dR}{dE_{e}} = \int_{T_{\chi}^{\min}}^{\infty} dT_{\chi} \int_{Q_{\min}}^{Q_{\max}} dQ \frac{dR}{dT_{\chi} dQ dE_{e}}.
\label{dR/dEe}
\end{equation}

The minimum kinetic energy \( T_{\chi}^{\min} \) can be derived using the kinematic limits in the relativistic case

\begin{equation}
T_{\chi}^{\min} = -m_{\chi} + \frac{E_{e}}{2} + \frac{1}{2} \sqrt{\frac{(2m_{e} + E_{e})(2m_{\chi}^{2} + m_{e}E_{e})}{m_{e}}}.
\label{kinetic limit with chi e}
\end{equation}

In semiconductor experiments, directly observing the distribution of $E_e$ is challenging. We thus focus on the average energy necessary to quantify the number of electron-hole pairs $n_e$ generated in a single event as a proxy for $E_e$. The observable event rate is given by

\begin{equation}
\frac{dR}{d n_{e}} = \int_{E_{e}^{\min}}^{\infty} dE_{e} \frac{dR}{dE_{e}} P(n_{e}, E_{e}),
\label{dR/dne}
\end{equation}
where \( P(n_{e}, E_{e}) \) is the probability that the electron-hole pair \( n_{e} \) is excited at \( E_{e} \). 
The data for $P(n_{e}, E_{e})$ are from \cite{Ramanathan:2020fwm} for a temperature of $100~\mathrm{Kelvin}$. Based on theoretical considerations related to \( Q \) and \( E_{e} \), the evaluation of the integrals ranges from a minimum value to infinity. It is worth noting that the contribution of the plasmon resonance is mainly confined to the very low range of electron transfer kinetic energies. Beyond these thresholds, significant suppression can be observed, and therefore we chose the integration ranges of \( \omega \in [1.11~\mathrm{eV}, 50~\mathrm{eV}] \) and \( Q \in [1.2~\mathrm{eV}, 5~\mathrm{keV}] \).

Fig.~\ref{Fig.4} illustrates the differential event rate within the detector for $m_{\chi}$ equaling $0.01~\mathrm{MeV}$, $0.1 ~\mathrm{MeV}$ and $1 ~\mathrm{MeV}$, where the left panel shows the differential event rate of the electron recoil energy and the right panel shows the one corresponding to the electron-hole pairs, with the selected benchmark cross-section $\bar{\sigma}_{\chi e}=10^{-34}~\mathrm{cm}^{2}$. Influenced by the energy loss function $\mathrm{Im}[-\epsilon(Q, E_{e})^{-1}]$, a peak in the differential event rate is observed at an electron energy $E_{e} \sim 15-20 \ \mathrm{eV}$, while the 3 to 6 electron-hole pairs dominate, that can be seen in the right panel 
of Fig.~\ref{Fig.4}. Furthermore, the differential event rate is proportional to $1/m_{\chi}^{3}$ for different $m_{\chi}$. The reason for such a difference is divided into two parts, where $1/m_{\chi}^{2}$ originates from the SN$\nu$BDM flux and $1/m_{\chi}$ from the DM-electron cross section.

For the SENSEI experiment at SNOLAB~\cite{SENSEI:2023zdf}, different response signals correspond to different electron-hole pairs, so the overall event number \( N_{\mathrm{Events}} \) must be multiplied by the corresponding effective exposures \( G_{n_{e}} \) for the various electron-hole pairs, as shown below:

\begin{equation}
N_{\mathrm{Events}} = \sum_{n_{e} = 3}^{10} G_{n_{e}} \frac{dR}{d n_{e}}.
\label{R}
\end{equation}

\begin{figure}[htb]
    \flushleft
    \includegraphics[width=0.48\textwidth,trim=3cm 0cm 3cm 0.5cm,clip]{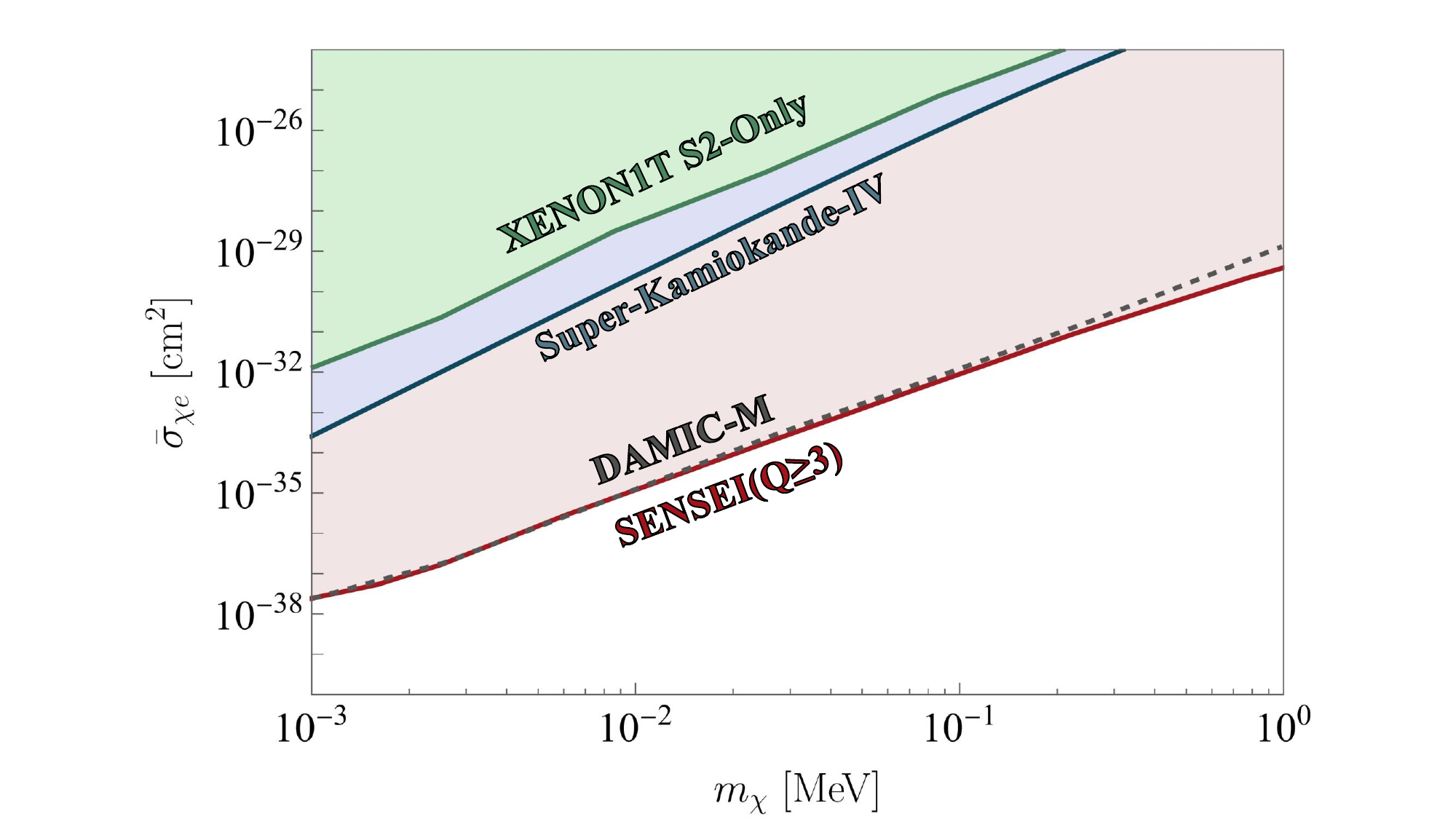}
\caption{Upper limit of the DM-electron cross section $\bar{\sigma}_{\chi e}$, where the red line corresponds to the four events from the SNOLAB SENSEI~\cite{SENSEI:2023zdf} experiment with threshold $\mathrm{Q} \geq 3$ and silicon detector exposure of $534.89\ \mathrm{g}\cdot \mathrm{day}$. The green line indicates constraints from the XENON1T experiment containing only S2 signals ~\cite{XENON:2019gfn}. The blue line is from the Super-Kamiokande-IV experiment constraint~\cite{Super-Kamiokande:2023jbt}, and the gray dashed line is from the DAMIC-M experiment constraint~\cite{DAMIC-M:2023gxo}.}
    \label{Fig.5}
\end{figure}

The total exposure of SENSEI at SNOLAB~\cite{SENSEI:2023zdf} is $534.89 \ \mathrm{g} \cdot \mathrm{days}$ for the silicon detectors. The effective exposure for each individual chamber can be retrieved from \cite{SENSEI:2023zdf}, where scenarios involving electron-hole pairs of three or more are considered, alongside a low background event rate. This approach is adopted to mitigate the impact of noise generated by $1 e^{-}$ and $2 e^{-}$ components on the detection process. Through likelihood data analysis, we can establish a limit on the SN$\nu$BDM cross-section, $\bar{\sigma}_{\chi e}$, with $g_{\nu} = g_{\chi} = 0.005$. The exclusion capability is illustrated by the red region in Fig.~\ref{Fig.5}.

As a competitive silicon semiconductor experiment, DAMIC-M has reported the SR2 searching data set with a corresponding exposure of $39.97~\mathrm{g}\cdot \mathrm{days}$~\cite{DAMIC-M:2023gxo}. Similar to the analysis of SENSEI, we select the data in the range of $3e^{-}$ to $5e^{-}$ to reduce the impact of background, which comprises 2 events. The exclusion capability of the DAMIC-M experiment is represented by the gray dashed line in Fig.~\ref{Fig.5}.

In contrast, Super-Kamiokande~\cite{Super-Kamiokande:2023jbt} relies on the Cherenkov radiation emitted from charged particles produced by neutrino interactions, with an energy threshold above \( 3.49~\mathrm{MeV} \). Thus, the electrons in the target can be considered free and at rest. It is generally believed that neutrino detectors have better sensitivity for boosted dark matter due to their significant exposure. However, we demonstrate in this paper that plasmon enhancement surpasses that of the neutrino detector. This is due to the exponential decline of the SN$\nu$BDM flux above \( 10 \, \mathrm{MeV} \), resulting in limited phase space for SN$\nu$BDM events above the Super-K threshold.

The differential event rate per unit target mass for Super-Kamiokande~\cite{Super-Kamiokande:2023jbt} is given by
\begin{equation}
	\frac{dR}{dE_e}
	=\mathcal{N}_e
	\int_{T_{\chi}^{\min}}dT_{\chi}
	\frac{d\sigma_{\chi e}}{dT_e} 
        \frac{d\phi_\chi}{dT_\chi},
	\label{eq:sk rate}
\end{equation}
where we assume the recoil electron scattered by DM follows the direction of the incoming DM particle. $\mathcal{N}_e$ is the electrons' number per unit target mass, e.g. for water Cherenkov detectors, $\mathcal{N}_e \approx 3.3 \times 10^{26}~\rm{kg^{-1}}$. $d\sigma_{\chi e}/dE_e$ is obtained from Eq.~\eqref{eq:cs nu DM} with subscript $\chi$ and $e$ interchanged.

We adopt the experimental data from Table C.1 in Ref.~\cite{Super-Kamiokande:2023jbt}, reported by the Super-Kamiokande~\cite{Super-Kamiokande:2023jbt} Collaboration, with an exposure time of 1664 days in a 22.5 kiloton fiducial target, resulting in an exposure of \( 102.6 \, \mathrm{kt \cdot yr} \). The observed number of events is 70,092 in the kinetic energy region of \( 3.49 \sim 19.5 \, \mathrm{MeV} \).
 
To constrain the dark matter cross section \( \bar{\sigma}_{\chi e} \) for different dark matter masses, we employ the conventional Poisson method. Assuming an expected event count of \( \lambda = s + b \), where \( s \) represents the theoretical number of signal events and \( b \) denotes the expected background, the probability of observing \( N_{\mathrm{obs}} \) events is given by the Poisson distribution:

\begin{equation}
P(N_{\mathrm{obs}} | \lambda) = \frac{e^{-\lambda} \lambda^{N_{\mathrm{obs}}}}{N_{\mathrm{obs}}!}
\end{equation}

The \( 1 - \alpha \) confidence level (C.L.) upper limit \( \lambda_p \) is calculated using the equation \( P(N < N_{\mathrm{obs}} | \lambda_p) = \alpha \). For instance, a \( 90\% \) C.L. corresponds to \( \alpha = 0.1 \). The exclusion region of the cross section \( \bar{\sigma}_{\chi e} \) is obtained by requiring \( \lambda < \lambda_p \).

In this work, the theoretical signal events $s$ is calculated by
\begin{equation}
   s= \epsilon \times E_{\rm x}\int dE_e \frac{dR}{dE_e},
\end{equation}
where the $\epsilon$ is the signal efficiency taken as 0.5 for conservative reasons, as the actual efficiency approximately between 0.5 and 0.75, $E_\mathrm{x}=102.6~\mathrm{kt\cdot yr}$ is the exposure of the Super-Kamiokande data~\cite{Super-Kamiokande:2023jbt}, The observed event $N_\mathrm{obs}=70092$. For the background event number,
we adopt the assumption that $b=N_{\mathrm{obs}}$ since the experiment does not detect any DM signals. The kinetic energy integral for the recoil electrons is from 3.49~MeV to 19.5~MeV. The corresponding exclusion region is represented by the blue region in Fig.~\ref{Fig.5}.

For completeness, we also include the XENON1T exclusion limit~\cite{XENON:2019gfn}. Since it is smaller than that of ~\cite{Super-Kamiokande:2023jbt}, we do not present its detailed derivation here. Interested readers can refer to \cite{Liang:2024xcx} for further details.

In summary, Fig.~\ref{Fig.5} illustrates the exclusion capability of SN$\nu$BDM for the DM-electron cross section. The red region represents the exclusion level achieved by the SENSEI experiment~\cite{SENSEI:2023zdf} as discussed before, the grey dashed line represents the exclusion level of the DAMIC-M experiment~\cite{DAMIC-M:2023gxo}, and the blue region shows the constraints imposed by the Super-Kamiokande-IV experiment~\cite{Super-Kamiokande:2023jbt}. The green region specifically denotes the limitations of the XENON1T experiment, considering only S2 signals~\cite{XENON:2019gfn}. Notably, in the case of a light mediator, semiconductor detectors exhibit a significant advantage over liquid detectors about $3$ to $4$ orders, primarily due to the plasmon enhancement effect in semiconductors. The exclusion limits obtained by SENSEI~\cite{SENSEI:2023zdf} and DAMIC-M~\cite{DAMIC-M:2023gxo}, both employing semiconductor detectors, exhibit consistency. This agreement demonstrates the strong competitiveness of both experiments in probing SN$\nu$BDM through the plasmon enhancement effect.

\section{Conclusions}
In this study, we investigate the phenomenon of SN$\nu$BDM and its implications for dark matter detection. We demonstrate that supernova neutrinos, emitted in copious amounts during core-collapse events, can transfer kinetic energy to dark matter particles, significantly enhancing their sensitivity. We derive an analytical expression for the spatial probability distribution of supernova neutrinos across various galactic locations, which is essential for considering dark matter-neutrino interactions. This aligns with existing literature on the limit of a single explosion and provides a comprehensive understanding of neutrino flux in the Milky Way. By calculating the resulting dark matter flux from supernova neutrinos, we establish a direct connection between neutrino interactions and dark matter detection and provide the inhomogeneous properties in angular distribution. Furthermore, we compute the plasmon excitation rate associated with SN$\nu$BDM, showing significant enhancements over traditional detection methods by approximately three orders, such as those used by Super-Kamiokande~\cite{Super-Kamiokande:2023jbt} and XENON1T~\cite{XENON:2019gfn}. The unique energy transfer mechanisms involved lead to a heightened sensitivity, particularly suitable for the solid-state detectors.

\section{Acknowledgement}
We thank Meng-Ru Wu and Liangliang Su for helpful discussions. This work is supported by the National Natural Science Foundation of China (NNSFC) under grant Nos. 12275232, 12275134, 12005180, by the Natural Science Foundation of Shandong Province under Grant No. ZR2020QA083, ZR2023QA149, and by the Project of Shandong Province Higher Educational Science and Technology Program under Grants No. 2022KJ271.

\bibliographystyle{JHEP}
\bibliography{refs}



%




\end{document}